\documentclass[12pt]{article}
\usepackage{geometry}                
\usepackage{graphicx}
\usepackage{amssymb}
\usepackage{epstopdf}
\usepackage{color}

\usepackage{amsmath}
\usepackage{graphics,graphicx,epsfig}
\usepackage{amssymb}
\usepackage{sidecap}
\sidecaptionvpos{figure}{c}
\usepackage{appendix}
\usepackage{cite}
\usepackage{xcolor}
\usepackage{mathtools}
\DeclareMathOperator{\sech}{sech}

\DeclareMathOperator{\arctanh}{arctanh}

\titlepage
\title{Static Spherically Symmetric Solutions in Conformally Coupled Scalar field}

\usepackage{geometry}                
\usepackage{graphicx}
\usepackage{amssymb}
\usepackage{epstopdf}
\usepackage{color}

\usepackage{amsmath}
\usepackage{graphics,graphicx,epsfig}
\usepackage{amssymb}
\usepackage{sidecap}
\sidecaptionvpos{figure}{c}
\usepackage{appendix}

\begin{document}

\begin{titlepage}
\vfill
\vskip 3.0cm
\begin{center}
{\Large\bf Revisiting the static spherically symmetric solutions of gravity with a conformally coupled scalar field}

\vskip 10.mm
{\bf  Sourya Ray} 
\vskip 0.4cm
Instituto de Ciencias F\'{\i}sicas y Matem\'{a}ticas, Universidad Austral de
Chile, Valdivia, Chile\\
\vskip 0.1 in Email: \texttt{ray@uach.cl}\\
\vspace{0.5 in}

\end{center}

\begin{center}
{\bf Abstract}
\end{center}

\begin{quote}
We revisit the static spherically symmetric solutions of Einstein's General Relativity with a conformally coupled scalar field in arbitrary dimensions. 
Using a four rank tensor introduced earlier we recast the field equations in a manifestly symmetric form to
elucidate a somewhat less-known feature of dual mapping between solutions.
We also show that there is a two-parameter subfamily of solutions which enjoy a duality symmetry and in four dimensions both the BBMB black hole and the Barcelo-Visser wormhole belong to this subfamily.
Along the way, we rederive the full three-parameter family of solutions by direct integration of the field equations and a natural choice of ansatz which arguably has several advantages over other previously known methods.
   \vfill
\vskip 2.mm
\end{quote}
\hfill
\end{titlepage}
\section{Introduction}

The energy-momentum tensor for a massless scalar field conformally coupled to gravity has long been known to have some interesting properties. It was shown in \cite{Callan:1970ze} to have finite matrix elements in every order of renormalized perturbation theory. Soon after Bekenstein showed that the conformally coupled scalar field system admits a black hole with a scalar charge thus providing one of the first counterexamples of the no-hair theorem \cite{Bekenstein:1974sf,Bekenstein:1975ts}. The solution was also independently obtained by Bocharova, Bronnikov and Melnikov in \cite{Bocharova:1970skc} and is commonly known as the BBMB black hole. It has the same geometry as that of an extreme Reissner-Nordstr\"om black hole although the scalar field becomes unbounded on the horizon. 

Growing interest in higher dimensional gravity led to the study of the conformally coupled scalar in higher dimensions. However, it was found that there are no higher dimensional analogues of the BBMB black hole. In fact a systematic search for black hole solutions among the general static spherically symmetric family resulted in the conclusion that the BBMB black hole is unique \cite{Xanthopoulos:1991} (see also \cite{Tomikawa:2016dqz,Tomikawa:2017vun}). Subsequently a two-parameter subfamily among the general class in four dimensions was identified to represent traversable wormholes connecting two asymptotically flat regions \cite{Barcelo:1999hq}.
The existence of the wormhole solution is attributed to the violation of the null energy condition.

It is often claimed that the field equations of this system are quite involved. So, a popular method to solve the system, discovered by Bekenstein himself, is to map the solutions of a minimally coupled scalar to that of a conformally coupled scalar. This map involves a conformal transformation followed by a suitable reparametrization of the scalar field. Although for static spherically symmetric solutions, direct integration of the field equations have been carried out \cite{Klimcik:1993cia}, they are usually, in our viewpoint, cumbersome and unintuitive.

In this short article, we reformulate the field equations in a manifestly symmetric form in order to simplify the field equations. This is done by using a four-rank tensor introduced previously which has the same algebraic symmetries as the Riemann tensor. We first demonstrate a duality relation between the solutions in arbitrary dimensions. The fact that the solutions of this system come in pairs was first observed by Bekenstein in four dimensions. We then solve the field equations for static spherically symmetric spacetimes by direct integration. To this end, we use an ansatz and an auxiliary scalar $\Psi$ which naturally exploits the symmetry of the equations and enables us to obtain the full three-parameter family of solutions in a straightforward manner. We then identify the duality relation between the solutions within this family and further point out a two-parameter subfamily which are self-dual. It is interesting to note that both the BBMB black hole and the Barcelo-Visser traversable wormhole (and its higher dimensional counterparts) belong to this subfamily, with the former as the limiting case of the latter. We also analyze the singularities of the quadratic curvature invariants for the whole family. Finally, we offer some further potential applications of our work.

\section{Conformally coupled scalar and duality of solutions}

In four spacetime dimensions, Einstein's General Relativity supplemented by a conformally coupled scalar field is usually given by the action
\begin{align}
I=\int \left[\left(1-\dfrac{1}{6}\kappa\phi^2\right)R+\kappa\phi\Box\phi\right] \sqrt{-g}\,d^4x
\end{align}
where $\kappa$ is a coupling constant. More generally, in arbitrary dimensions $D$, it takes the form
\begin{align}
I=\int \left[\left(1-\dfrac{(D-2)}{4(D-1)}\kappa\phi^2\right)R+\kappa\phi\Box\phi\right]\sqrt{-g}\,d^Dx
\end{align}
However, the action and the corresponding field equations can be recast in a more symmetrical manner by introducing the following four-rank tensor which has the same algebraic symmetries as the Riemann tensor \cite{Oliva:2011np}
\begin{align}\label{DefS}
S_{\alpha\beta}^{\ \ \mu\nu}=\phi^2R_{\alpha\beta}^{\ \ \mu\nu}+\dfrac{4}{s}\phi\delta_{[\alpha}^{[\mu}\phi_{;\beta]}^{\ \nu]}+\dfrac{4(1-s)}{s^2}\delta_{[\alpha}^{[\mu}\phi_{;\beta]}\phi_{;}^{\ \nu]}-\dfrac{2}{s^2}\delta_{[\alpha}^{[\mu}\delta_{\beta]}^{\nu]}\phi_{;\lambda}\phi_{;}^{\ \lambda}
\end{align}
where $\phi_{;\alpha}=\nabla_{\alpha}\phi$ and $\phi_{;\alpha}^{\ \beta}=g^{\beta\mu}\nabla_{\alpha}\nabla_{\mu}\phi$. Let us also note that this tensor can be obtained from the Riemann tensor by a conformal rescaling of the metric. More precisely,
\begin{align}\label{RSrel}
S_{\alpha\beta}^{\ \ \mu\nu}=\phi^{\frac{2(s-1)}{s}}R_{\alpha\beta}^{\ \ \mu\nu}[\phi^{-\frac{2}{s}}g_{\lambda\sigma}]
\end{align}
Then, under the conformal transformation
\begin{align*}
g_{\mu\nu}\rightarrow \Omega^2g_{\mu\nu}, \qquad \phi\rightarrow \Omega^s\phi
\end{align*}
this tensor transforms as
\begin{align*}
S_{\alpha\beta}^{\ \ \mu\nu}\rightarrow \Omega^{2(s-1)}S_{\alpha\beta}^{\ \ \mu\nu}.
\end{align*}

The real quantity $s$ characterizes the conformal weight of the scalar field and can be chosen arbitrarily except $0$. We also define the contractions
\begin{align*}
S_{\alpha}^{\ \beta}=S_{\alpha\mu}^{\ \ \beta\mu}\qquad \& \qquad S=S_{\alpha}^{\ \alpha}
\end{align*}
which are analogously related to the Ricci tensor and scalar curvature through a conformal rescaling.
In terms of this tensor, the action is compactly expressed as
\begin{align*}
I=\int \left[R-\kappa\phi^mS\right]\sqrt{-g}\,d^Dx
\end{align*}
where $m=(2(1-s)-D)/s$. Furthermore, we can set $m=0$ by taking $s=1-D/2$ without any loss of generality. This essentially amounts to a (conformal) rescaling of the scalar field $\phi$. This particular choice reduces the action to the more familiar form given earlier.

Varying the resulting action with respect to the metric gives
\begin{align*}
G_{\alpha}^{\ \beta}=\kappa\left[S_{\alpha}^{\ \beta}-\dfrac{1}{2}S\delta_{\alpha}^{\beta}\right]
\end{align*}
whereas variation with respect to the scalar field gives the conformally invariant equation $S=0$, which in turn, along with the trace of the equation above, leads to $R=0$. Thus, the equations of motion can be summarized as
\begin{align}\label{feq1}
&\qquad R_{\alpha}^{\ \beta}=\kappa S_{\alpha}^{\ \beta}\\
&\label{feq2}S=0 \quad \Leftrightarrow \quad R=0.
\end{align}
Finally, before we confine ourselves to static solutions, let us point out another of the advantages of our formulation. Suppose we write $g_{\mu\nu}=\Omega^2\tilde{g}_{\mu\nu}$, where $\Omega$ is any real function. Then, due to the property (\ref{RSrel}), the field equation (\ref{feq1}) becomes
\begin{align*}
R_{\alpha}^{\ \beta}[\Omega^2\tilde{g}_{\mu\nu}]=\kappa S_{\alpha}^{\ \beta}=\kappa \phi^{\frac{2D}{D-2}}R_{\alpha}^{\ \beta}[\Omega^2\phi^{\frac{4}{D-2}}\tilde{g}_{\mu\nu}]
\end{align*} 
We next introduce an auxiliary field $\Psi=\Omega \phi^{2/(D-2)}$ \footnote{For arbitrary $s$, $\Psi=\Omega\phi^{-1/s}$. The rest of the equations are then independent of $s$.}. After multiplying both sides by $\Omega^D$ and setting $\kappa=1$, we get
\begin{align}\label{Rfeq1}
\Omega^DR_{\alpha}^{\ \beta}[\Omega^2\tilde{g}_{\mu\nu}]-(\Omega\rightarrow\Psi)=0.
\end{align}
where the second term inside parenthesis can simply be obtained from the first by replacing $\Omega$ with $\Psi$. The equation for the scalar field can be expressed as
\begin{align}\label{Rfeq2}
R[\Psi^2\tilde{g}_{\mu\nu}]=0 \quad \Leftrightarrow \quad R[\Omega^2\tilde{g}_{\mu\nu}]=0.
\end{align} 
Note that the equations (\ref{Rfeq1}) and (\ref{Rfeq2}) are manifestly symmetric under $\Omega\leftrightarrow\Psi$ interchange. Explicitly, this implies that for every solution $(g_{\mu\nu},\phi)$ to the system, there is a dual solution $(\phi^{4/(D-2)}g_{\mu\nu},\phi^{-1})$. This duality in four dimensions was first observed by Bekenstein (Theorem 2 in \cite{Bekenstein:1974sf}) and more recently in \cite{Martinez:2020hjm} in higher dimensions. Here we clarify the underlying origin of the duality explicitly without any reference to solutions to the minimally coupled scalar field system.

There is a class of trivial solutions to the equations (\ref{Rfeq1}) and (\ref{Rfeq2}) for which the scalar field $\phi$ takes a constant value $\phi_0$ everywhere. In this case the equations reduce to
\begin{align*}
(1-\phi_0^2)\Omega^DR_{\alpha}^{\ \beta}[\Omega^2\tilde{g}_{\mu\nu}]=0 \quad\text{and}\quad R[\Omega^2\tilde{g}_{\mu\nu}]=0
\end{align*} 
If $\phi_0\neq \pm1$, then any Ricci-flat metric solves the field equations. However, if $\phi_0=\pm 1$, then any metric with $R=0$ is a solution. We shall exclude these trivial solutions from further discussions below.

\section{Static spherically symmetric solution}

We now analyze the static spherically symmetric solutions of the field equations (\ref{feq1}) and (\ref{feq2}). These solutions were previously obtained in \cite{Xanthopoulos:1992fm} using transformations mapping solutions from the minimally coupled scalar field to those coupled conformally and in \cite{Klimcik:1993cia} via direct integration of the field equations. In either case, the solutions were obtained in isotropic coordinates where two unknown metric functions are solved for.
Here, we will show that the reformulation of the conformal coupling in terms of the $S$-tensor naturally motivates an alternate ansatz which not only elucidates several aspects of the phase space of the solutions but also allows a simpler and a more systematic direct integration of the field equations (See also \cite{Martinez:2020hjm}).

Considering the fact that the $S$-tensor is related to the Riemann tensor by a conformal rescaling, it is then natural to take one of the unknown functions in the metric ansatz to be an overall conformal factor. The other unknown function can be chosen in multiple equivalent ways, here we choose it to be the $g_{tt}$ component of the conformally related metric i.e., our ansatz is
\begin{align}
ds^2=\Omega(r)^2\left[-f(r)^2\,dt^2+dr^2+d\Sigma_{D-2}^2\right]
\end{align}
where $d\Sigma_{D-2}^2=\tilde{g}_{ij}d\tilde{x}^id\tilde{x}^j$ is the line element of a $(D-2)-$dimensional sphere. The Ricci tensor components are then given by
\begin{align}
&R_t^{\ t}=-\dfrac{\Omega^{-D}f^{-1}}{D-2}\partial_r(f^{-(D-3)}\partial_r(\Omega f)^{D-2})\\
&R_r^{\ r}=-\Omega^{-2}\left[\partial_r^2\ln|f\Omega^{D-1}|+(\partial_r\ln|f|)(\partial_r\ln|\Omega f|)\right]\\
&R_j^{\ i}=-\dfrac{\Omega^{-D}}{(D-2)}\left[-(D-3)(D-2)\Omega^{D-2}+f^{-1}\partial_r(f\partial_r\Omega^{D-2})\right]\delta_j^i
\end{align}
while the scalar curvature is given by
\begin{align}
R&=-\Omega^{-2}\left[-(D-3)(D-2)+2\partial_r^2\ln|f\Omega^{D-1}|+(D-2)(D-1)(\partial_r\ln|\Omega|)^2\right.\nonumber\\
&\left.\:\:\:+2(D-1)(\partial_r\ln|\Omega|)(\partial_r\ln|f|)+2(\partial_r\ln|f|)^2\right]
\end{align}
We first solve the following combination of the scalar equations
\begin{align}
\Omega^2R-(\Omega\rightarrow\Psi)=\dfrac{2(D-1)}{f(\Omega\Psi)^{\frac{D-2}{2}}}\partial_r(f(\Omega\Psi)^{\frac{D-2}{2}}\partial_r\ln|\Omega/\Psi|)=0
\end{align}
The above equation suggests the following reparametrizations of $\Omega$ and $\Psi$ in terms of $z(r)$ and $\omega(r)$
\begin{align}
\Omega^{\frac{D-2}{2}}=e^{\left(\frac{D-2}{D-3}\right)z}\cosh\omega\qquad \& \qquad \Psi^{\frac{D-2}{2}}=e^{\left(\frac{D-2}{D-3}\right)z}\sinh\omega
\end{align}
which brings the solution to the form
\begin{align}\label{1stint}
fe^{\frac{2(D-2)z}{D-3}}\partial_r\omega=C_4
\end{align}
If $C_4=0$, then $\omega=$ constant which in turn implies that the scalar field $\phi=(\Psi/\Omega)^{\frac{D-2}{2}}=\phi_0$ is constant everywhere and the solution belongs to the trivial class discussed earlier.
Henceforth, we shall consider $C_4\neq 0$. We then solve the $^t_t$ -component of the field equations (\ref{Rfeq1}) which gives
\begin{align}
f^{-(D-3)}\partial_r(f^{D-2}e^{\frac{2(D-2)z}{D-3}})=2(D-2)C_4C_2
\end{align}
which upon using (\ref{1stint}) becomes
\begin{align}
\partial_{\omega}\ln|fe^{\frac{2z}{D-3}}|=2C_2
\end{align}
and can be integrated to obtain
\begin{align}\label{2ndint}
fe^{\frac{2(D-2)z}{D-3}}=\pm e^{2\tilde{z}}=C_4\partial_\omega r
\end{align}
where $\tilde{z}=z+C_2\omega+C_3$. Next, we solve the $^i_j$ -component of the field equations (\ref{Rfeq1}) giving
\begin{align*}
f^{-1}\partial_r(f\partial_re^{\frac{2(D-2)z}{D-3}})=(D-3)(D-2)e^{\frac{2(D-2)z}{D-3}}
\end{align*}
which on using (\ref{2ndint}) reduces to
\begin{align}
\partial_{\omega}^2\tilde{z}=\dfrac{(D-3)^2}{2C_4^2}e^{4\tilde{z}}
\end{align}
and can be solved using elementary methods noting that the right hand side does not depend on the variable $\omega$ explicitly. The solution is given by
\begin{align*}
&e^{4\tilde{z}}=\dfrac{4C_4^2C_1^2}{(D-3)^2}\sinh^2[(D-3)(r-r_0)]\\
\quad\text{and}\quad &\tanh[(D-3)(r-r_0)/2]=\pm\exp[2C_1(\omega-\omega_0)]
\end{align*}
The remaining field equations are solved provided the integration constants $C_1$ and $C_2$ satisfy the relation
\begin{align}\label{constraint}
C_1^2-C_2^2=\dfrac{(D-1)(D-3)}{(D-2)^2}
\end{align}
The integration constant $r_0$ can be set to zero by a translation of the coordinate $r$. Similarly, the constant $C_4(\neq 0)$ can be chosen arbitrarily by rescaling the time coordinate $t$. Hence, we obtain a 3-parameter family of solutions characterized by $C_1$ (or $C_2$), $C_3$ and $\omega_0$. Plugging everything back and expressing in terms of the rescaled coordinate $\tilde{r}=(D-3)r/2$ we obtain
\begin{align}\label{gensol1}
&f(r)^2=\left(\sech\tilde{r}\right)^{\frac{4}{D-3}}\left(\tanh\tilde{r}\right)^{\frac{2((D-2)C_2-C_1)}{(D-3)C_1}}\\
&\Omega(r)^{D-2}-\Psi(r)^{D-2}=A\left(\cosh\tilde{r}\right)^{2\left(\frac{D-2}{D-3}\right)}\left(\tanh\tilde{r}\right)^{\frac{(D-2)(C_1-C_2)}{(D-3)C_1}}
\end{align}
while
\begin{align}\label{gensol2}
\phi(r)=\left(\dfrac{\Psi(r)}{\Omega(r)}\right)^{\frac{D-2}{2}}=\pm\dfrac{(\tanh\tilde{r})^{\frac{1}{C_1}}-B}{(\tanh\tilde{r})^{\frac{1}{C_1}}+B}
\end{align}
where the constants $C_3$ and $\omega_0$ have been traded for $A(\neq 0)$ and $B(\neq0,\infty)$ respectively.

Notice that for a fixed $\{C_1,C_2\}$ the mapping $\{A,B\}\to\{-A,-B\}$ corresponds to a mapping to the dual solution i.e., it is equivalent to interchanging $\Omega$ and $\Psi$.

We reiterate the simplicity of the direct integration of the field equations in this formulation and the natural choice of our ansatz in contrast to the ones existing in the literature. In the following sections we demonstrate some further advantages.

\section{Self-dual solutions}
As was explained earlier, since the field equations are invariant under $\Omega \leftrightarrow \Psi$ interchange, there is a dual mapping between solutions. In general, the dual solutions correspond to distinct points in the parameter space and are physically distinct. However, we show below that there exist special points in the parameter space of the solutions for which the dual solutions are simply related by a (discontinuous) coordinate transformation and hence are self-dual. 

Among the non-trivial class, the condition for self duality is $(D-2)C_2=C_1=\pm1$. Imposing this condition the solution reduces to
\begin{align*}
&f(r)^2=(\sech\tilde{r})^{\frac{4}{D-3}};\\
&\Omega(r)^{D-2}-\Psi(r)^{D-2}=A\left(\tanh\tilde{r}\right)(\cosh\tilde{r})^{2\left(\frac{D-2}{D-3}\right)}\\
&\phi(r)=\left(\dfrac{\Psi(r)}{\Omega(r)}\right)^{\frac{D-2}{2}}=\pm\dfrac{b-e^{-2\tilde{r}}}{1-be^{-2\tilde{r}}}=\pm\left(\dfrac{b-e^{2\tilde{r}}}{1-be^{2\tilde{r}}}\right)^{-1}
\end{align*}
where $B$ in the previous section has been replaced by $(1-b)/(1+b)$, where $b$ now takes on any real values except $\pm1$. It is evident that for the above two-parameter family, the solutions related by $\Omega\leftrightarrow\Psi$ interchange simply corresponds to the transformation $r\leftrightarrow -r$. 

The metric can also be easily expressed in the isotropic coordinates by making the transformations
\begin{align*}
{\bar{r}}^{D-3}=\mu e^{2\tilde{r}}\qquad \text{and} \qquad {\bar{t}}^{D-3}=4\mu t^{D-3}
\end{align*}
and redefining $A=(1-b^2)(4\mu)^{(D-2)/(D-3)}$, which brings the line element to
\begin{align*}
ds^2=\left(1-\dfrac{b\mu}{\bar{r}^{D-3}}\right)^{\frac{4}{D-2}}\left[-\left(1+\dfrac{\mu}{\bar{r}^{D-3}}\right)^{-\frac{4}{D-2}}d\bar{t}^2+\left(1+\dfrac{\mu}{\bar{r}^{D-3}}\right)^{\frac{4}{(D-2)(D-3)}}(d\bar{r}^2+\bar{r}^2d\Sigma_{D-2}^2)\right]
\end{align*}
which is manifestly asymptotically flat while the scalar field is given by
\begin{align*}
\phi(r)=\pm\dfrac{b\bar{r}^{D-3}-\mu}{\bar{r}^{D-3}-b\mu}
\end{align*}
It is interesting to note that the two well-known physically relevant spacetimes belong to this subclass of solutions. In $D=4$, this solution was shown in \cite{Barcelo:1999hq} to represent a traversable wormhole for negative values of $b$, 
where the constants are related by $\mu=\eta/2$ and $b=\tan{\Delta/2}$.
While for $b=0$ in $D\geq4$ the above family of asymptotically flat self-dual solutions were termed Bekenstein spacetimes in \cite{Klimcik:1993cia}, which contains the BBMB black hole in $D=4$. 

\section{Singularities, horizons and throats}

In this section we examine some physical properties of the solutions. We begin by performing a simple analysis of the curvature singularities. Since the scalar curvature vanishes, there are only two independent squared curvature invariants: $C_{\alpha\beta}^{\ \ \rho\sigma}C^{\ \ \alpha\beta}_{\rho\sigma}$ and $R_{\alpha\beta}R^{\alpha\beta}$ whose singularities we analyze below.
\begin{enumerate}
\item $\mathbf{C_{\alpha\beta}^{\ \ \rho\sigma}C^{\ \ \alpha\beta}_{\rho\sigma}}$:
Due to spherical symmetry every non-zero component of the conformal (Weyl) tensor are proportional to each other while the conformal factor only appears as a multiplicative prefactor. Explicitly,
\begin{align*}
C_{\alpha\beta}^{\ \ \rho\sigma}C^{\ \ \alpha\beta}_{\rho\sigma}=\dfrac{4(D-3)}{(D-1)}\Omega(r)^{-4}\left(\dfrac{f''(r)}{f(r)}-1\right)^2
\end{align*} 
We first evaluate the term inside the parenthesis. A little algebra shows
\begin{align*}
&\dfrac{f''(r)}{f(r)}-1=\dfrac{(D-2)}{4C_1^2}\left[\left((D-2)C_2-C_1\right)\left(C_2-C_1\right)\coth^2\tilde{r}\right.\nonumber\\
&\qquad\qquad\qquad\left.+\left((D-2)C_2+C_1\right)\left(C_2+C_1\right)\tanh^2\tilde{r}-2((D-2)C_2^2+C_1^2)\right]
\end{align*}
So unless the coefficient of the hyperbolic cotangent (squared) on the right hand side vanishes, as $r\to 0$
\begin{align}\label{C^2omega^4asrgoesto0}
\left(\dfrac{f''(r)}{f(r)}-1\right)^2\sim\coth^4\tilde{r}
\end{align}
Next, we examine the behavior of the conformal factor $\Omega$ near $r=0$.
Using the expressions in (\ref{gensol1}) and (\ref{gensol2}) we find that as $r\to 0$, 
\begin{align}\label{omegasrgoesto0}
\Omega(r)^{-4}\sim\tanh^\alpha\tilde{r}
\end{align}
where
\begin{align*}
\alpha=4\left(\dfrac{1}{(D-2)|C_1|}-\dfrac{1-C_2/C_1}{D-3}\right)<\dfrac{4}{(D-2)|C_1|}\leqslant\dfrac{4}{\sqrt{(D-3)(D-1)}}
\end{align*}
where we have used the constraint (\ref{constraint}). Hence, $\alpha<4$ for $D\geqslant4$. Combining (\ref{C^2omega^4asrgoesto0}) and (\ref{omegasrgoesto0}) we see that the conformal invariant is singular at $r=0$ for any dimensions $D \geqslant 4$ unless $(D-2)C_2=C_1$ which is precisely the condition for self-duality of the solutions. Therefore, \textit{all the non-self-dual solutions have a naked singularity at $r=0$}.

We now look at the singularities of the two-parameter family of self-dual solutions, in which case
\begin{align*}
C_{\alpha\beta}^{\ \ \rho\sigma}C^{\ \ \alpha\beta}_{\rho\sigma}=(D-3)(D-1)\left(\dfrac{4B}{A}\right)^{4/(D-2)}(\tanh\tilde{r}+B)^{\frac{-8}{(D-2)}}\left(\sech\tilde{r}\right)^{\frac{4(D-1)}{D-3}}
\end{align*}
\textit{For $|B|\geqslant 1$ the above invariant is finite for all values of $r$ and in particular it vanishes as $r \to \pm\infty$, while for $|B|<1$ it is singular at $\tilde{r}=\arctanh(-B)$ since the conformal factor $\Omega(r)$ blows up there}.

\item $\mathbf{R_{\alpha\beta}R^{\alpha\beta}}$: We look for singularities in this invariant only for the self-dual family. The non-zero components of the Ricci tensor for this family are
\begin{align*}
R^{\ t}_t&=-\dfrac{D-3}{2}\left(\dfrac{D-3}{D-2}\right)\dfrac{A}{4B}\Omega(r)^{-D}(\sech{\tilde{r}})^{\frac{2(D-4)}{D-3}}\\
R^{\ r}_r&=-\dfrac{1}{2}\left(\dfrac{D-3}{D-2}\right)\dfrac{A}{4B}\Omega(r)^{-D}(\sech{\tilde{r}})^{\frac{2(D-4)}{D-3}}\left[1-(D-2)^2(1-B^2)\cosh^2\tilde{r}\right]\\
R^{\ i}_j&=\delta^i_j\dfrac{1}{2}\left(\dfrac{D-3}{D-2}\right)\dfrac{A}{4B}\Omega(r)^{-D}(\sech{\tilde{r}})^{\frac{2(D-4)}{D-3}}\left[1-(D-2)(1-B^2)\cosh^2\tilde{r}\right]
\end{align*}
where
\begin{align}\label{conformalfactor}
\Omega(r)^{D-2}=\dfrac{A}{4B}(\tanh\tilde{r}+B)^{2}\left(\cosh\tilde{r}\right)^{\frac{2(D-2)}{D-3}}
\end{align}

First note that \textit{for $|B|<1$, exactly as the conformal tensor, each component of the Ricci tensor above is singular at $\tilde{r}=\arctanh(-B)$ due to the singularity of the conformal factor $\Omega(r)$}. 

Next, for $|B|=1$ all the components are proportional to the function
\begin{align*}
\Omega(r)^{-D}(\sech{\tilde{r}})^{\frac{2(D-4)}{D-3}}=\exp{\left(\mp\frac{2D\tilde{r}}{D-2}\right)}(\sech\tilde{r})^{\frac{4(D-2)}{D-3}-\frac{2D}{D-2}}
\end{align*}
Since, as $\tilde{r}\to\pm\infty$, $\sech\tilde{r}\to2e^{-|\tilde{r}|}$, the expression above is non-singular at $r=\pm\infty$ only if
\begin{align*}
\dfrac{4(D-2)}{D-3}-\dfrac{4D}{D-2}=-\dfrac{4(D-4)}{(D-2)(D-3)}\geqslant0
\end{align*}
which holds only for $D=4$. Hence, \textit{for $|B|=1$, the invariant $R_{\alpha\beta}R^{\alpha\beta}$ is singular at $r=\infty$ or $-\infty$ except for $D=4$ in which case it reduces to $\Omega(r)^{-8}$ up to a proportionality constant while $\Omega(r)$ is finite at either of the two locations and becomes infinite at the other depending on the sign of $B$}. The point where $\Omega(r)$ is finite represents a horizon since $f(r)$ vanishes there and the metric can be analytically extended beyond this point by a coordinate transformation while the other represents an asymptotic region.
This exceptional four dimensional spacetime represents the well-known BBMB black hole and has the same geometry as the extreme Reissner-Nordstr\"om black hole. 

Finally, \textit{for $|B|>1$ each component is regular for the whole range $-\infty<r<\infty$ and in particular vanishes at both the boundaries and hence so is the curvature invariant $R_{\alpha\beta}R^{\alpha\beta}$}. It is also not too difficult to show from (\ref{conformalfactor}) that in this case $\Omega(r)\to\infty$ as $r\to\pm\infty$ and has a minimum somewhere in between. In order to see the latter, we equate the derivative $\Omega'(r)$ to $0$ and obtain the following quadratic equation in $\tanh\tilde{r}$
\begin{align*}
P(\tanh\tilde{r})\coloneqq\tanh^2\tilde{r}+(D-2)B\tanh\tilde{r}+(D-3)=0
\end{align*}
which for $|B|>1$ has two real roots $\xi_1,\xi_2$. Since $\xi_1\xi_2=(D-3)>0$ while $\xi_1+\xi_2=-(D-2)B$, the two roots have the same sign and opposite to that of $B$. Suppose $|\xi_1|<|\xi_2|$. Moreover, $P(\pm1)=(D-2)(1\pm B)$. Hence,
\begin{align*}
\text{If}\: B>1,\: \text{then}\: P(-1)<0 \implies \xi_2<-1<\xi_1<0\\
\text{If}\: B<-1,\: \text{then}\: P(1)<0 \implies 0<\xi_1<1<\xi_2
\end{align*}
i.e., $\xi_1$ lies between $0$ and $\pm1$ depending on whether $B$ is positive or negative while $|\xi_2|>1$. Therefore, $\Omega(r)$ has exactly one minima at $\tanh\tilde{r}_m=\xi_1$ where
\begin{align*}
\xi_1=-\dfrac{(D-2)B}{2}\left(1-\sqrt{1-\dfrac{4(D-3)}{(D-2)^2B^2}}\right)
\end{align*}
The spacetime represents the well-known asymptotically flat `traversable' wormhole \cite{Barcelo:1999hq} obtained by Barcelo and Visser and its higher dimensional generalization with $\tilde{r}=\tilde{r}_m$ as the wormhole throat\footnote{To the best of our knowledge, the higher dimensional generalization of the Barcelo-Visser wormhole has not been pointed out in the literature previously.}. The scalar field $\phi$ approaches mutually reciprocal (finite) values at the two asymptotic regions.

\end{enumerate} 

\section{Conclusions}

Recasting the field equations for a scalar field conformally coupled to gravity in a manifestly symmetrical form, we have shed light on a duality feature among the solutions. Next, focusing on the static spherically symmetric solutions we have identified a two-parameter subfamily which are self-dual and are related by a point reflection transformation. In four spacetime dimensions this subfamily is shown to include both the BBMB black hole and the Barcelo-Visser `traversable' wormhole. In the process, we have provided a direct integration method to obtain the full three-parameter family of solutions in any dimension $D\geqslant4$ in a simpler way than existing in the literature.



The duality among solutions is generally broken in presence of other  matter fields or even in presence of a cosmological constant \cite{Ortaggio:2024afv}. It is also broken if the gravity theory is arbitrarily modified. However, if both the gravity part and the matter coupling part are modified in the `same way' then the duality symmetry will continue to exist. 
Note that the tensor $S_{\alpha\beta}^{\mu\nu}$ used here was originally introduced in \cite{Oliva:2011np} in order to construct higher curvature analogues of conformal coupling and interpreted as a scalar field conformally coupled to Lovelock gravity which is the higher curvature generalization of General Relativity.
So, if a Lovelock theory of a fixed order (also known as `pure' Lovelock theory) is appended by the corresponding higher curvature conformal coupling, then the corresponding equations of motion will exhibit the duality symmetry. It would be interesting to see if the symmetry can be similarly exploited in such a theory to integrate the field equations and obtain exact solutions.


Although, here we have confined ourselves to static spherically symmetric spacetimes, as explained in section 2, the duality among solutions is valid for any solution and could be useful to study other exact solutions. 
For example, an analytical solution representing a self-similar collapse of a spherically symmetric conformally coupled scalar field in four spacetime dimensions is known \cite{deOliveira:1995cn}. More recently, a class of stationary axisymmetric solutions have been found using solution generating techniques which contains a rotating generalization of the BBMB black hole \cite{Astorino:2014mda}. It is worth classifying these solutions based on the duality relation. Work along these lines is in progress.

\textbf{Acknowledgements:} The author gratefully acknowledges many useful discussions with Marcello Ortaggio and Hideki Maeda.

\end{document}